\documentclass[12pt]{article}
\usepackage{graphicx,amsfonts,amsmath,amssymb,mathrsfs}

\title{On Superselection Rules in Bohm--Bell Theories}

\author{ 
Samuel Colin\footnote{Perimeter Institute for Theoretical Physics, 
    31 Caroline Street North, Waterloo, Ontario, Canada, N2L 2Y5.
    E-mail: scolin@perimeterinstitute.ca},
Thomas Durt\footnote{Fakulteit der Wetenschappen, Vrije 
   Universiteit Brussel, Pleinlaan 2, 1050 Brussel, Belgium.
   E-mail: thomdurt@vub.ac.be}, and
Roderich Tumulka\footnote{Mathematisches Institut,
    Eberhard-Karls-Unversit\"at, Auf der Morgenstelle 10, 72076
    T\"ubingen, Germany.  E-mail:
    tumulka@everest.mathematik.uni-tuebingen.de}
}
\date{November 7, 2006}

\newcommand{\Hilbert}{\mathscr{H}}
\newcommand{\conf}{\mathcal{Q}}
\newcommand{\Q}{\conf}
\newcommand{\tr}{\mathrm{tr}}

\renewcommand{\Re}{\mathrm{Re}}
\renewcommand{\Im}{\mathrm{Im}}
\newcommand{\EEE}{\mathbb{E}}
\newcommand{\PPP}{\mathbb{P}}
\newcommand{\RRR}{\mathbb{R}}
\newcommand{\CCC}{\mathbb{C}}

\newcommand{\prob}{\mathrm{Prob}}
\newcommand{\sphere}{\mathbb{S}} 
\renewcommand{\sp}[2]{\langle #1|#2 \rangle}
\newcommand{\Laplace}{\Delta} %
\newcommand{\pvm}{P}%
\newcommand{\obs}{G} 
\newcommand{\fct}{g} 
\newcommand{\ev}{\gamma} 
\newcommand{\fer}{\mathrm{f}} 
\newcommand{\bos}{\mathrm{b}} 
\newcommand{\vk}{\boldsymbol{k}}
\newcommand{\vx}{\boldsymbol{x}}
\newcommand{\vy}{\boldsymbol{y}}
\newcommand{\vQ}{\boldsymbol{Q}}
\newcommand{\sys}{\mathsf{sys}} 
\newcommand{\env}{\mathsf{env}} 
\newcommand{\inter}{\mathsf{int}} 
\newcommand{\spin}{\mathrm{spin}} 
\newcommand{\posi}{\mathrm{pos}} 
\newcommand{\histories}{\mathcal{Z}} 
\newcommand{\Endproof}{\hfill $\square$}

\newtheorem{prop}{Proposition}

\newcommand{\y}[1]{{#1}}
\newcommand{\z}[1]{{#1}}

\begin{document}
\maketitle
\hyphenation{super-selection}
\begin{abstract}
The meaning of superselection rules in Bohm--Bell theories (i.e., quantum theories with particle trajectories) is different from that in orthodox quantum theory. More precisely, there are two concepts of superselection rule, a weak and a strong one. Weak superselection rules exist both in orthodox quantum theory and in Bohm--Bell theories and represent the conventional understanding of superselection rules. We introduce the concept of strong superselection rule, which does not exist in orthodox quantum theory. It relies on the clear ontology of Bohm--Bell theories and is a sharper and, in the Bohm--Bell context, more fundamental notion. A strong superselection rule for the observable $\obs$ asserts that one can replace every state vector by a suitable statistical mixture of eigenvectors of $\obs$ without changing the particle trajectories or their probabilities. A weak superselection rule asserts that every state vector is empirically indistinguishable from a suitable statistical mixture of eigenvectors of $\obs$. We establish conditions on $\obs$ for both kinds of superselection. For comparison, we also consider both kinds of superselection in theories of spontaneous wave function collapse.

\medskip

\end{abstract}

\section{Introduction}

Bohm--Bell theories are quantum theories with particle trajectories guided by the quantum state vector $|\psi \rangle$ in such a way that at every time $t$ the configuration $Q_t$ has probability distribution $|\psi_t|^2$. We give a detailed definition in Section~\ref{sec:cond}. Among these theories are Bohmian mechanics \cite{Bohm52,DGZ92,survey,Gol01} (a theory for nonrelativistic quantum mechanics), Bell's jump process for lattice quantum field theory \cite{Bell86,crex1}, and ``Bell-type quantum field theories'' \cite{crlet, crea2B, samuelthesis} (the generalization of both Bohmian mechanics and Bell's process to quantum field theory in the continuum). These theories were introduced for solving the conceptual difficulties of quantum theory \cite{Gol01}.

Recall that a \emph{superselection rule} for a quantum observable $\obs$ is, naively speaking, the statement that $\obs$ always ``assumes a sharp value'', i.e., that wave functions are always eigenfunctions of $\obs$, whereas nontrivial (coherent) superpositions of eigenfunctions of $\obs$ with different eigenvalues do not occur in nature. Since the \emph{superposition principle} of quantum theory asserts that for any two quantum states, also their complex linear combinations are quantum states, a more careful formulation of a superselection rule asserts that one can always replace, without loss of generality, a wave function by a statistical (incoherent) mixture of eigenfunctions of $\obs$. But what exactly does that mean here, ``without loss of generality''? There are two possible answers, and thus two interpretations of superselection rules: a strong, unconventional one that does not refer to observers and is available in Bohm--Bell theories but not in orthodox quantum mechanics; and a weak, conventional one that refers to observers and is available in both theories.

Of the two interpretations of superselection rules, we begin with describing the weak one. It asserts that no possible experiment can distinguish between the actual wave function, which could be a nontrivial superposition of eigenfunctions of $\obs$ with different eigenvalues, and a suitable statistical mixture of eigenfunctions of $\obs$ \cite{WWW52, Wi95, BW03}. This amounts to the statement that not all self-adjoint operators on the Hilbert space of a system correspond to observables, or, in other words, that some operators (those which do not commute with $\obs$) cannot ``be measured.'' A difficulty with arguing for a weak superselection rule in any particular case is that no criterion is known for which operators do correspond to executable experiments and which do not; this may render the justification of a superselection rule uncertain and unsatisfactory. On top of that, in orthodox quantum theory, where operators as observables are introduced \emph{by postulate}, any reasoning concerning which operators correspond to observables and which do not, is likely to have the ring of arbitrariness. In total, it is not clear on which principles a claim of weak superselection should be based. In this respect, one is better off in Bohm--Bell theories, in which the connection between experiments and operators is \emph{derived} rather than postulated. In Section~\ref{sec:unmeasurable} we will illustrate this with a concrete example and formulate natural conditions on $\obs$ for a weak superselection rule.

But now we turn to the strong interpretation of the superselection rule for $\obs$ that is available in Bohm--Bell theories. It asserts that whatever the wave function $\psi$, there is a mixture $\mu^\psi$ of eigenfunctions $\phi$ of $\obs$ that leads to precisely the same trajectories of all particles at all times as $\psi$ with the same probabilities. That is, not only are the outcomes of all experiments the same, but even all microscopic facts about the path of every single particle. To appreciate that this is genuinely more, note that in a Bohmian universe there are strong limitations on the access of macroscopic observers to the details of microscopic trajectories \cite{DGZ92}, because to observe means to influence. Since for a superselection rule in this sense it is also true that the state vector is empirically indistinguishable from a suitable mixture of eigenvectors of $\obs$, strong superselection is indeed stronger than weak superselection. 

Here is an example, concerning a simple Bell-type quantum field theory discussed in \cite{crea2B,crea1}. The model, whose defining equations are given in Section~\ref{sec:ex}, involves two species of particles, one fermionic and one bosonic, such that the fermions can emit and absorb bosons. A strong superselection rule holds in this model for the fermion number operator, as the trajectories of the particles, as well as the emission and absorption rates, depend only on the part of the state vector $\psi$ in the appropriate superselection sector, i.e., in the eigenspace of the fermion number operator whose eigenvalue coincides with the number of fermions in the actual configuration. Thus, on the level of the particle trajectories there is no difference between a superposition and a mixture of different fermion numbers. For more detail about this example see Section~\ref{sec:ex}.

We now define strong superselection with mathematical precision. We start with a Hilbert space $\Hilbert$ on which $\obs$ is an operator; $\Hilbert$ contains, of course, all superpositions of eigenvectors of $\obs$ with different eigenvalues. For the unit sphere in $\Hilbert$ we write $\sphere(\Hilbert) = \{\psi \in \Hilbert: \|\psi\| = 1\}$. A statistical mixture of wave functions is mathematically described by a probability measure $\mu$ on $\sphere(\Hilbert)$.\footnote{The reader might think that a statistical mixture is mathematically described by a density matrix. However, while the density matrix encodes all information relevant to the statistics of outcomes of experiments, it does not contain enough information for the Bohmian trajectories and their statistics, as first emphasized by Bell \cite{Belldm}.} Let $\obs$ be a self-adjoint operator with pure point spectrum (otherwise its eigenvectors would not span $\Hilbert$) and $E_\ev$ the eigenspace of $\obs$ with eigenvalue $\ev$. Let $\Q$ be the configuration space in which the configuration $Q_t$ moves, and $H$ the Hamiltonian. (A typical example of the configuration space in a Bohm--Bell theory is the set of all finite subsets of $\RRR^3$, denoted $\Gamma_{\neq} (\RRR^3) = \{Q \subseteq \RRR^3: \# Q < \infty\}$, whose elements represent the positions of a variable, finite number of identical particles. \z{Generally, we will assume that $\Q$ is a countable union of disjoint manifolds.}) \z{Abstractly}, the motion (and, possibly, creation and annihilation) of the particles is mathematically described by a stochastic process $Q=(Q_t)_{t\in \RRR}$ in $\Q$ that depends on the initial state vector $\psi \in \sphere(\Hilbert)$, $Q = Q^\psi$; the process $Q^\psi$ is characterized by its distribution $\PPP^\psi$, a probability measure on the path space $\histories$ of $\Q$; $\histories$ is a space of mappings $\RRR \to \Q$ from the time axis to the configuration space, representing the possible histories. For Bohmian mechanics, e.g., $\histories$ is the space of continuous curves in $\Q$, and for Bohm--Bell theories in general \cite{crea2B}, due to the possibility of jumps, it is the space of piecewise continuous curves in $\Q$.

We define that \emph{a strong superselection rule holds for $\obs$ if for every $\psi \in \sphere(\Hilbert)$ there is a mixture $\mu^\psi$ concentrated on the eigenvectors of $\obs$ such that}
\begin{equation}\label{sssr}
  \int\limits_{\phi \in \sphere(\Hilbert) \cap \cup_\ev E_\ev} 
  \mu^\psi(d\phi) \, \PPP^\phi(dQ) = \PPP^\psi(dQ)\,. 
\end{equation}
In words, the same trajectories with the same probabilities are generated by a mixture of eigenvectors $\phi$ of $\obs$ (with distribution $\mu^\psi$) as by $\psi$. \y{(Since the symbol $\PPP$ was sometimes \cite{crlet,crea2B} used for the distribution at time $t$ in \emph{configuration space}, we emphasize again that here, $\PPP^\psi$ denotes the distribution in \emph{path space}.)}

The remainder of this paper is organized as follows. In Section~\ref{sec:cond} we formulate simple conditions for strong superselection. In Section~\ref{sec:ex} we describe examples of strong superselection. In Section~\ref{sec:det} we mention a link between strong superselection and determinism. In Section~\ref{sec:unmeasurable} we discuss weak superselection, including examples and conditions for it. In Section~\ref{sec:grw} we discuss superselection in theories of spontaneous wave function collapse. In Section~\ref{sec:conclusions} we conclude. Proofs are collected in the Appendix.

\section{Conditions for Strong Superselection}
\label{sec:cond}

We can formulate natural sufficient (and presumably also necessary) conditions for a strong superselection rule after defining more precisely what we mean by a Bohm--Bell theory.

A Bohm--Bell theory can be defined from the following structure as data \cite{crea2B}: a Hilbert space $\Hilbert$, a Hamiltonian $H$ on $\Hilbert$, a state vector $|\psi\rangle \in \Hilbert$ that evolves according to the Schr\"odinger equation
\begin{equation}\label{schr}
  i\hbar \frac{d|\psi_t \rangle}{dt} = H|\psi_t \rangle \,,
\end{equation}
a configuration space $\Q$, and a projection-valued measure (PVM) $\pvm(dq)$ on $\Q$ acting on $\Hilbert$ that serves as the configuration observable, i.e., the totality of all position observables. The particle configuration follows a stochastic or deterministic process $(Q_t)_{t \in \RRR}$ in $\Q$ that can be defined as follows. The evolution of $Q_t$ consists of continuous motion interrupted by stochastic jumps. Let the Hamiltonian $H$ possess the decomposition $H = H_0 + H_I$ into a differential operator $H_0$ (often the free Hamiltonian) and an integral operator $H_I$ (often the interaction Hamiltonian). We assume that the operator $H_0$ is, at every configuration $q$, either of the Schr\"odinger type,
\begin{equation}\label{H0Schr}
  H_0 = \sum_{i,j} a_{ij}(q) \frac{\partial^2}{\partial q_i \partial q_j} 
  + \sum_i b_i(q) \frac{\partial}{\partial q_i} + V(q)
\end{equation}
with positive definite matrices $a_{ij}(q)=a_{ji}(q)$, or of the Dirac type,
\begin{equation}\label{H0Dirac}
  H_0 = \sum_i b_i(q) \frac{\partial}{\partial q_i} + V(q), 
\end{equation}
where $b_i(q)$ are (in every space direction) matrices of full rank on spin space (or, generally speaking, on the value space of the wave function). The continuous motion of $Q_t$ is determined by \cite{crea2B}
\begin{equation}\label{bohm}
  \frac{dQ_t}{dt} = v^{\psi_t} (Q_t) \text{ with } v^\psi \cdot 
  \nabla f(q) = \Re \frac{\sp{\psi} { \pvm(dq) \tfrac{i}{\hbar} 
  [H_0, F] |\psi }}{\sp{\psi| \pvm(dq)}{\psi} } \:\: \forall f \in C_0^\infty(\Q)\,,
\end{equation}
where $C_0^\infty(\Q)$ denotes the space of all smooth functions $f: \Q \to \RRR$ with compact support, and the operator $F$ is the function $f$ of the configuration observable $\pvm(dq)$, which means that
\begin{equation}\label{Fdef}
  F = \int_\Q f(q) \,\pvm(dq) \,.
\end{equation}
(On an $L^2$ space with its natural PVM, this is the multiplication operator by the function $f$.) The continuous motion is interrupted by stochastic jumps $q' \to q$ that occur with rate 
\begin{equation}\label{bell}
  \sigma^{\psi_t} (dq|q') = \frac{ [\tfrac{2}{\hbar} \, \Im \, \sp{\psi_t}
  { \pvm(dq) H_I \pvm(dq') | \psi_t}]^+} {\sp{\psi_t} {\pvm(dq') | \psi_t}} \,,
\end{equation}
where $x^+ = \max(x,0)$. For a detailed discussion of this process, see \cite{crea2B}.

\begin{prop}
Let $\obs$ be a self-adjoint operator with pure point spectrum. In Bohm--Bell theories as defined above, a strong superselection rule holds for $\obs$ if
\begin{subequations}\label{conditions}
\begin{align}
  \obs \text{ is a function } \fct: \Q \to \RRR & \text{ of 
  the configuration observable } \pvm(dq), \text{ and} \label{fctcondition}\\
  & [\obs,H] =0 \label{commcondition}\,.
\end{align}
\end{subequations}
\end{prop}

(All proofs are postponed to the Appendix.)
Furthermore and more explicitly, under the conditions \eqref{conditions} we have:
\begin{itemize}
\item[(a)] $[\obs,H_0]=0$ and $[\obs,H_I]=0$.

\item[(b)] The value $\ev = \fct(Q_t)$ is an eigenvalue of $\obs$, as, in fact, the only values that the function $\fct$ can assume are the eigenvalues of $\obs$. 

\item[(c)] The function $\fct$ is constant on every connected component of $\Q$, as a consequence of $[\obs,H_0] = 0$. (The \emph{connected components} of the set $\Q$ are defined by the property that two points lie in the same connected component whenever there is a continuous path from one to the other. For example, the connected components of $\Gamma_{\neq} (\RRR^3)$ are the $n$-particle sectors for $n= 0,1,2,\ldots$, where the $n$-particle sector is the set of all $n$-element subsets of $\RRR^3$, $\Gamma_n (\RRR^3) = \{Q \subseteq \RRR^3: \# Q = n\}$.)

\item[(d)] With probability one, $\fct(Q_t)$ is a conserved quantity, i.e., this value is time-independent, or constant along the trajectory $t \mapsto Q_t$. \z{This is a trivial consequence of (c) if the trajectory is continuous, but it holds as well for the stochastic jumps obeying \eqref{bell}.} Since $\fct(Q_t)$ can be regarded as ``the actual value of the observable $\obs$,'' we have conservation laws on both levels, that of operators and that of the actual configuration.

\item[(e)] Parts of the state vector $\psi_t$ are irrelevant for the time evolution of the configuration $Q_t$; indeed, 
\begin{equation}\label{Pevpsi}
  \text{the evolution of }Q_t \text{ depends only on }
  \pvm_\obs(\ev) \psi_t \,,
\end{equation}
where $\pvm_\obs(\ev)$ denotes the projection to $E_{\ev}$, the eigenspace of $\obs$ corresponding to the eigenvalue $\ev$.  We use the notation 
\begin{equation}
  \psi^\ev := \frac{\pvm_\obs(\ev) \psi}{ \|\pvm_\obs(\ev) \psi\|}
\end{equation}
for the (renormalized) component with eigenvalue $\ev$. Note that, by \eqref{commcondition}, $\pvm_\obs(\ev) \psi_t = (\pvm_\obs(\ev) \psi)_t$ and thus $(\psi_t)^\ev = (\psi^\ev)_t$. If the evolution of $Q_t$ is deterministic, a case in which $Q_t$ is a function of the initial state vector $\psi_0$ and the initial configuration $Q_0$, $Q_t = Q_t(\psi_0,Q_0)$, \eqref{Pevpsi} means that $Q_t = Q_t(\psi_0^\ev, Q_0)$.

\item[(f)] Also the probability distribution of $Q_t$ (conditional on the value $\ev$) is unchanged, i.e.,
\begin{equation}\label{probpsiev}
  \PPP^{\psi^\ev}\Bigl(Q_t \in dq \Bigr) = \PPP^\psi \Bigl(Q_t 
  \in dq \Big| \fct(Q_t) = \ev \Bigr)\,.
\end{equation}
In fact, if $\psi$ can be written as a function on configuration space then the left hand side is $|\psi_t^\ev(q)|^2 \, dq$ and the right hand side is, using the notation $1_B(q)$ for the indicator function of the set $B$ which is $1$ for $q\in B$ and $0$ otherwise,
\[
  \| \pvm_\obs(\ev) \psi_t\|^{-2} \, 1_{\{\fct = \ev\}}(q) \, |\psi_t(q)|^2 \, dq\,.
\]

\item[(g)] The mixture $\mu^\psi$ consists of
\begin{equation}\label{mixture}
  \text{the state vectors } \psi^\ev 
  \text{ with probabilities } \|\pvm_\obs(\ev) \psi\|^2 \,,
\end{equation}
and has density matrix
\begin{equation}\label{dm}
  \rho^\psi = \sum_{\ev} \pvm_\obs (\ev) |\psi\rangle \langle \psi| \pvm_\obs(\ev) \,.
\end{equation}
\end{itemize}

\begin{prop}\label{twoiffeight}
A set of conditions equivalent to \eqref{conditions} is
\begin{subequations}\label{altconditions}
\begin{align}
  &\obs \text{ is a function } \fct: \Q \to \RRR \text{ of 
  the configuration observable } \pvm(dq).\\
  &\text{With probability one, } \fct(Q_t) \text{ is a conserved quantity.} \label{conservcondition}
\end{align}
\end{subequations}
\end{prop}

Since we could not think of any counterexample, we conjecture that the conditions \eqref{conditions}, respectively \eqref{altconditions}, are not only sufficient but also necessary for strong superselection.

\section{Examples}
\label{sec:ex}

Let us consider an explicit example: a simple Bell-type quantum field theory discussed in detail in \cite{crea2B}, with two particle species, one fermionic and one bosonic; for simplicity, both species are scalar (spin zero). The Hilbert space $\Hilbert$ is the tensor product of a fermionic and a bosonic Fock space, and the Hamiltonian is given by
\begin{equation}\label{Hex}
\begin{split}
  H = &\int d^3 \vk \, \omega_\fer (\vk) \, a_\fer^*(\vk) 
  \, a_\fer(\vk) + \int d^3 \vk \, \omega_\bos (\vk) 
  \, a_\bos^*(\vk) \, a_\bos(\vk) \: +\\
  +& \int d^3 \vx \, a_\fer^*(\vx) \biggl(
  \int d^3 \vy \, \varphi(\vx-\vy) \bigl( a_\bos^*(\vy) + a_\bos(\vy) 
  \bigr) \biggr) a_\fer(\vx)\,.
\end{split}
\end{equation}
Here $a^*$ and $a$ are the creation and annihilation operators, either for the fermions or for the bosons depending to the subscript  and either in the momentum representation or in the position representation depending on the argument $\vk$ or $\vx,\vy$; $\omega(\vk)$ is the dispersion relation (either for the fermions or for the bosons depending on the subscript), which we take, for simplicity, to be the nonrelativistic one, $\omega_i(\vk) = \hbar^2 \vk^2/2m_i$, $i=\fer,\bos$; and $\varphi(\vx)$ is a continuous function strongly peaked at the origin that serves for regularizing the Hamiltonian. Then $H_0$ is the sum of the first two integrals of \eqref{Hex}, and $H_I$ is the third. The configuration space is 
\begin{equation}\label{Qex}
  \Q = \Gamma_{\neq} (\RRR^3) \times \Gamma_{\neq} (\RRR^3)\,, 
\end{equation}
where the two factors in the Cartesian product correspond to the fermions and the bosons, respectively, and the symbol $\Gamma_{\neq}$ was defined in Section 1. \y{As a consequence of $\omega_i(\vk) = \hbar^2 \vk^2/2m_i$, $H_0$ is in the position representation a differential operator of the Schr\"odinger type \eqref{H0Schr}; indeed, at a configuration $q$ with $N_\fer$ fermions and $N_\bos$ bosons, $H_0$ acts as \cite{crea1}
\begin{equation}
  - \frac{\hbar^2}{2m_\fer} \sum_{k=1}^{N_\fer} \Laplace_{\fer,k} - 
  \frac{\hbar^2}{2m_\bos} \sum_{k=1}^{N_\bos} \Laplace_{\bos,k}\,,
\end{equation}
with $\Laplace_{\fer,k}$ and $\Laplace_{\bos,k}$ the Laplacians in the $k$-th fermion/boson coordinate.} 

The laws governing the particles in this model are given by the equations \eqref{bohm} and \eqref{bell}. In this model, finitely many particles move in $\RRR^3$, each of them either a fermion or a boson, and every fermion can emit a boson, thus increasing the number of particles by one. The emission event occurs spontaneously, that is, stochastically with a rate given in terms of the state vector. Furthermore, every fermion can spontaneously absorb a boson when it has come close enough.

A strong superselection rule holds here for the \emph{number of fermions}, corresponding to $\fct(Q) = \fct(Q_\fer, Q_\bos) = \# Q_\fer$. \z{Condition} \eqref{commcondition} is easy to check. Indeed, the fermion number is conserved, as it changes neither at boson emission or absorption nor during the mere motion of particles. As a consequence, it is only the sector of Hilbert space corresponding to the actual fermion number that is relevant to the behavior of the particles. In more detail, we can decompose the Hilbert space into \z{particle-number} sectors,
\begin{equation}
  \Hilbert = \bigoplus_{N_\fer=0}^\infty \bigoplus_{N_\bos =0}^\infty 
  \Hilbert^{(N_\fer,N_\bos)} \,,
\end{equation}
where $\Hilbert^{(N_\fer,N_\bos)}$ is the space of states with $N_\fer$ fermions and $N_\bos$ bosons; of the parts $\psi^{(N_\fer,N_\bos)}$ of the state vector $\psi$ that lie in $\Hilbert^{(N_\fer,N_\bos)}$, only those with $N_\fer = \# Q_\fer$ govern the behavior of the particles. (Indeed, when this model was first described in \cite{crea1}, the remainder of the Hilbert space was left out right from the start, taking instead $\Hilbert = \oplus_{N_\bos =0}^\infty \Hilbert^{(N_\fer,N_\bos)}$ with $N_\fer$ a fixed number.)

Another example is provided by a Bell-type version of a simple quantum field theory described in detail in \cite{crea2B}, involving a second-quantized Dirac field in an external electromagnetic field and featuring electron--positron pair creation and annihilation. \z{The configuration space is again of the form \eqref{Qex}, with the first factor now corresponding to electrons and the second to positrons.} In this model, a finite number of particles move in $\RRR^3$, and each of the particles is identified as either an electron or a positron. (Suitable conditions are assumed of the external field that ensure that the number of particles stays finite for all times.) Creation events, at which the number of particles increases by two (one electron and one positron), occur spontaneously, that is, stochastically with rates determined by the state vector and the external field. Similarly, annihilation events can occur whenever an electron and a positron are sufficiently close to each other. A strong superselection rule holds here for the \emph{total charge}, i.e., the number of positrons minus the number of electrons. 

A similar example is provided by a Bell-type version of quantum electrodynamics outlined in \cite{SC,samuelthesis}, involving infinitely many particles (electrons of positive or negative energy). In this model, the Dirac sea is taken literally, so that what is usually regarded as the vacuum state is associated with infinitely many electrons of negative energy (with actual positions); pair creation, in contrast, is not to be taken literally but corresponds to the excitation of a negative energy electron to positive energy. In fact, in this model no particle is ever created or annihilated, and consequently, the total particle number, or, equivalently, the total charge, is conserved (after subtracting an infinite constant). Thus, the total number or total charge operator (after subtracting an infinite constant) is strongly superselected.

As an example of an operator that satisfies \eqref{commcondition} but not \eqref{fctcondition}, and in fact is not strongly superselected, consider, in Bohmian mechanics with $\Q = \RRR^d$, $\Hilbert = L^2(\Q)$, $H_I = 0$, and $H_0 = -\tfrac{\hbar^2}{2m} \nabla^2 +V$ with $V(-q) = V(q)$, the parity operator
\begin{equation}
  \obs \psi(q) = \psi(-q)
\end{equation}
whose eigenspaces are the even functions ($\ev=1$) and the odd functions ($\ev=-1$). Indeed, $\obs$ is not strongly superselected because the particle velocities generically depend both on the even and the odd part of the wave function.

It may be useful to have an example of strong superselection involving only Bohmian mechanics. The primary examples of superselection rules, of course, arise from quantum field theory, and not from $N$-particle quantum mechanics, and that is why the following example from quantum mechanics is slightly artificial. We consider Bohmian mechanics in a 3-space (a Riemannian 3-manifold) $C_1 \cup C_2$ with two connected components $C_1$ and $C_2$, corresponding to $\Q = (C_1 \cup C_2)^N$. This situation can be thought of as arising in the following two ways. We can first regard this as an effective description in the presence of an infinitely high potential barrier separating the regions $C_1$ and $C_2$ of $\RRR^3$. Alternatively, suppose the geometry of space-time on a cosmological level were such that, in a suitable space + time splitting, 3-space evolves from approximately a 3-sphere to approximately two 3-spheres; that is, space-time has an upside-down ``pair of pants'' topology. Then, from some time onwards, 3-space has two connected components, $C_1$ and $C_2$. Let $\obs$ be the number of the component in which particle 1 is, corresponding to $\fct(q) = \fct(q_1, \ldots, q_N) = 1_{\{q_1 \in C_1\}} + 2 \cdot 1_{\{q_1 \in C_2\}}$. Then $\obs$ is strongly superselected, as it satisfies \eqref{altconditions}.

\section{Determinism}
\label{sec:det}

There is a link between superselection rules and determinism: in a deterministic Bohm--Bell theory, the total number of particles is always strongly superselected. 

To begin with, a Bohm--Bell theory is deterministic if and only if all jump rates vanish, $\sigma^\psi(dq|q') = 0$ for all $\psi$. The following proposition tells us when this happens.

\begin{prop}\label{propdet}
A Bohm--Bell theory is deterministic if and only if $H_I$ commutes with the configuration observable,
\begin{equation}
  [H_I, \pvm(B)] =0 \quad \forall B \subseteq \Q\,.
\end{equation}
\end{prop}

Since in a Bohm--Bell theory, every point in configuration space corresponds to a particle configuration, the total number of particles is a function $\fct$ of the configuration and is constant on every connected component of configuration space. In a deterministic Bohm--Bell theory, since $Q_t$ moves continuously, the total number of particles $\fct(Q_t)$ is conserved, satisfying \eqref{conservcondition}. Thus, the corresponding operator $\obs$ is strongly superselected, \z{and indeed $[\obs,H_0]=0=[\obs,H_I]$}.

An example of a deterministic model is the Bell-type quantum electrodynamics with infinitely many electrons \cite{SC,samuelthesis} that we have already mentioned, with strongly superselected charge operator.

\section{Weak Superselection}
\label{sec:unmeasurable}

\z{The concept of weak superselection is based on what we can macroscopically observe, and thus inherits the fuzziness associated with the notions ``we'', ``macroscopic'', and ``observe''. Still, and in a way surprisingly, we can formulate precise conditions sufficient for weak superselection, and prove them in the context of Bohm--Bell theories. We begin with an example.}

Which operators on the Hilbert space of a system correspond to experiments that we can perform on the system depends on which interactions we can arrange between the system and the apparatus. Here is a concrete example of such a limitation: spin could not be ``measured'' if there were no magnetic fields. 

As an explicit model in the framework of Bohm--Bell theories, consider a nonrelativistic world with $N$ spin-$s$ particles in $\RRR^3$ in which the only potentials are Coulomb potentials: that is, $\Hilbert = L^2(\RRR^{3N}, (\CCC^{2s+1})^{\otimes N})$ with the natural configuration observable corresponding to the configuration space $\Q = \RRR^{3N}$, and
\begin{equation}\label{exH}
  H = -\sum_{i=1}^N \frac{\hbar^2}{2m_i} \nabla_i^2 + \sum_{i<j} 
  \frac{e_i \, e_j} {|\vx_i - \vx_j|}
\end{equation}
with $m_i$ and $e_i$ the mass and the charge of the $i$-th particle. The particle trajectories are given by Bohmian mechanics, with law of motion 
\begin{equation}\label{exbohm}
  \frac{d\vQ_i}{dt} = \frac{\hbar}{m_i} \Im \frac
  {\psi^* \nabla_i\psi}{\psi^* \psi} (\vQ_1, \ldots, \vQ_N)\,,
\end{equation}
where $\phi^*\psi$ denotes the inner product in spin space $(\CCC^{2s+1})^{\otimes N}$. \y{Eq.~\eqref{exbohm}} is the special case of \eqref{bohm} with $H$ given by \eqref{exH} and the natural PVM on $\RRR^{3N}$. In such a world, all observables that one can ``measure'' on a system of $n$ particles act trivially on the spin degrees of freedom, i.e., they are of the form $A = A_\posi \otimes 1_\spin$, where $A_\posi$ acts on $L^2(\RRR^{3N},\CCC)$ (the Hilbert space of the position degrees of freedom), and $1_\spin$ is the identity on $(\CCC^{2s+1})^{\otimes N}$ (the spin space). To see this, we may exchange the ``up'' and ``down'' components of one particle (a procedure corresponding to a unitary operator $U = 1_\posi \otimes U_\spin$ on $\Hilbert$) and observe in \eqref{exbohm} that $U\psi$ leads to the same trajectories (and the same probabilities) as $\psi$ since permutation of spin components does not change the velocity and $U$ commutes with $H$. Therefore, it is impossible to distinguish the two spin components on the basis of any information about the particle trajectories, and thus impossible to perform a ``spin measurement.''\footnote{To be sure, a wave function that is a superposition of ``up'' and ``down'' can lead to trajectories that would not arise from either a pure spin-up or a pure spin-down wave function; that is why this example is not an example of a strong superselection rule.} As a consequence, a weak superselection rule holds for every spin matrix: superpositions of spin eigenstates are empirically indistinguishable from statistical mixtures of spin eigenstates.

The abstract structure of this example is as follows. Consider a model world with Hilbert space $\Hilbert$ and Hamiltonian $H$, and a subsystem ``$\sys$'' on which its environment ``$\env$'' performs experiments. Mathematically, let $\Hilbert = \Hilbert_\sys \otimes \Hilbert_\env$ and 
\begin{equation}\label{Hcompo}
  H = H_\sys \otimes 1_\env + 1_\sys \otimes H_\env + H_\inter\,, 
\end{equation}
where $H_\inter$ is the interaction between ``$\sys$'' and ``$\env$.'' (Note that \z{the operator} $\obs$ need not be an \emph{observable}. Indeed, in the example above, $\obs$ is a Pauli spin matrix.)

\begin{prop}\label{propweak}
Let $\obs$ be a self-adjoint operator on $\Hilbert_\sys$ with pure point spectrum. In Bohm--Bell theories, a weak superselection rule holds for $\obs$ if
\begin{equation}\label{condwssr}
  [\obs,H_\sys] = 0 = [\obs \otimes 1_\env,H_\inter]  
\end{equation}
\end{prop}

Here is another, alternative, criterion, \z{which does not presuppose a division into system and environment.}

\begin{prop}\label{GHPweak}
Let $\obs$ be a self-adjoint operator on $\Hilbert$ with pure point spectrum. In Bohm--Bell theories, a weak superselection rule holds for $\obs$ if
\begin{subequations}\label{GHPwssr}
\begin{align}
  [\obs, \pvm(B)] &=0 \quad \forall B \subseteq \Q \,,\label{GcommP}\\
  [\obs,H] &=0\,.\label{GcommH}
\end{align}
\end{subequations}
\end{prop}

Let us compare the two criteria. It may seem surprising that no commutation between $\obs$ and the configuration observable is required in Proposition~\ref{propweak}: after all, a weakly superselected operator would be expected to commute with all observables (even though this does not, \z{perhaps,} strictly follow from the definition). The answer is that, as a consequence of the commutation with $H_\inter$, in practice $\obs$ does commute with $\pvm(dq)$, or at least with the ``macroscopic configuration observable'', obtained by suitably coarse-graining $\pvm(dq)$. (In the latter case, the name ``configuration observable'' for $\pvm(dq)$ would be not quite appropriate because it would not be fully observable.) To be sure, there are mathematical examples of operators $H_\inter$ (such as $H_\inter =0$) for which not even the macroscopic configuration is observable, but that does not happen in practice.

Which operators are observable and which are not, though according to the orthodox spirit it may have to be postulated, comes out of an analysis of the interaction between the system and its environment. This is exactly what we will do in the proof of Proposition~\ref{propweak}, but the same analysis can be done without Bohmian mechanics if one is willing to accept a certain gap in the analysis, corresponding to the quantum measurement problem. This attitude lies somewhere between orthodox quantum mechanics and Bohm--Bell theories and is typical of the ``decoherence'' approach.

A surprising trait of Proposition~\ref{GHPweak} is that it does not require a splitting of the world into system and environment (or apparatus or observer); indeed, $\obs$ can be an operator on the entire world, e.g., the total charge of the universe. The orthodox formalism, in contrast, always assumes such a splitting, with the funny consequence that an observer cannot measure, \z{e.g.,} her own body weight. Thus, the condition \eqref{GHPwssr} cannot be derived from orthodox quantum theory, but can from Bohm--Bell theories where no such difficulty arises. This circumstance is particularly relevant since the prime examples of superselection rules concern the entire universe, such as total charge or total baryon number. To be sure, we have not found the condition \eqref{condwssr} in the literature either, but we expect it may well exist somewhere.

It could be that \eqref{GcommP}, given \eqref{GcommH}, is not merely sufficient but also necessary for weak superselection. (In fact, if \eqref{GcommP} is violated then at least the obvious mixture \eqref{mixture} leads to a different distribution of the configuration, which should in principle be observable.) If this is the case then every weak superselection rule with \eqref{GcommH} that has been empirically obtained or confirmed (such as, e.g., the charge of the universe, or its baryon number) restricts the possible choices of the configuration observable $\pvm(dq)$. That gives us a way of selecting $\pvm(dq)$ in cases in which different choices are possible.

Proposition~\ref{GHPweak} expresses how the Hamiltonian and the position operators determine weak superselection rules. Combining it with Noether's theorem, we obtain as a corollary that every continuous symmetry that leaves both the Hamiltonian and the position operators invariant is generated by a weakly superselected operator. Examples of this situation are gauge symmetries, replacing the field operator $\Psi(\vx)$ by $e^{i\theta}\Psi(\vx)$, which give rise to the weak superselection of the corresponding charge operator.

\y{It is worth noting a major difference between the (weak or strong)
superselection rules we are dealing with and the so-called
``environment-induced superselection'' rules \cite{Giulini}, which is a more approximate concept: While environment-induced superselection makes it \emph{difficult} to see 
interference between different sectors, weak superselection makes it
\emph{impossible}.}

Let us turn to another example, similar to the world without magnetic fields considered in the beginning of this section: a world in which magnetic fields can point only in the $z$ direction. Then $\sigma_z$, the $z$ component of spin, can ``be measured'', but no other spin component can. Furthermore, $\sigma_z$ is weakly superselected (by either Proposition~\ref{propweak} or \ref{GHPweak}), but no other spin component is. Moreover, $\sigma_z$ is not a function of the configuration observable, and indeed it is not strongly superselected since the velocities of the particles generically depend both on the spin-up and the spin-down amplitude of the wave function. For comparison, in the model without magnetic fields, no spin component can ``be measured'', every spin component is weakly and none strongly superselected.

Note that a physicist living in the example world with all magnetic fields along the $z$ axis may conjecture from her experiences that the spin-up and spin-down parts of the wave function correspond to two distinct species of particles, which can be expressed mathematically by taking as the configuration space, instead of $\RRR^{3N} = (\RRR^{3})^N$, 
\begin{equation}
  \Q = (\{\mathrm{up},  \mathrm{down}\} \times \RRR^3)^N 
  = \Q_\spin \times \RRR^{3N}\,,
\end{equation}
where $\Q_\spin = \{\mathrm{up}, \mathrm{down}\}^N$ is a discrete set with $2^N$ elements. The physicist would be led to a different Bohmian theory, physically different though empirically indistinguishable from the one with $\Q = \RRR^{3N}$, in which the velocity of a particle depends only on the part of the wave function corresponding to its ``actual spin''. In this Bohmian theory, the operator $\sigma_z$ is, in fact, strongly superselected. This example illustrates that it may depend on the choice of the configuration observable $\pvm$ and the configuration space $\Q$ (and thus on the ontology) whether a given \y{operator} $\obs$ is strongly superselected or not.

\section{GRW}
\label{sec:grw}

Another approach besides Bohm--Bell theories providing quantum theories without observers is based on the assumption of spontaneous collapses of the wave function \cite{BG03}. The best-known model of this kind is due to Ghirardi, Rimini, and Weber (GRW) \cite{GRW86, Bell87a}, designed for nonrelativistic $N$-particle quantum mechanics. This model has much in common with Bohm--Bell theories \cite{AGTZ}, which makes it interesting to compare the status of superselection rules. In particular, the GRW model possesses a notion of strong superselection, whose relevance, however, depends on the choice of \emph{primitive ontology} \cite{AGTZ}, i.e., of what should be regarded as the constituents of reality that, say, tables and chairs are made of. The primitive ontology of Bohm--Bell theories, for example, is formed by the particle trajectories. The GRW model allows several choices of primitive ontology. With the ``flash'' ontology, all examples of superselection we know are examples of strong superselection, whereas with the ``matter density'' ontology, all examples we know are examples of weak superselection.

The version ``GRWf'' \cite{Bell87a,AGTZ} is based on the flash ontology. In this version, the primitive ontology is formed by discrete space-time points called ``flashes'' (the centers of the wave function collapses), and the path space we used in Bohm--Bell theories is replaced by the space $\histories$ of all $N$-tuples $(S_1, \ldots, S_N)$ of discrete subsets of space-time, $S_i$ being the set of all flashes associated with particle number $i$. The history of a GRWf world corresponds to one element of $\histories$, chosen at random according to some probability measure $\PPP^\psi$ on $\histories$ depending on the initial wave function $\psi$. In this abstract terminology, the definition of strong superselection around eq.~\eqref{sssr} can be adopted without change. And indeed, a strong superselection rule holds in essentially the same cases as in Bohmian mechanics; we now formulate a sufficient condition.

Following \cite{Tu05}, the mathematical structure of a GRW-type theory with flash ontology is defined in terms of a Hamiltonian $H$; the flash rate operators $\Lambda(\vx), \vx \in \RRR^3$, a family of positive operators (in the original GRW model, $\Lambda(\vx)$ is the multiplication operator by a Gaussian centered at $\vx$); and a vector $\psi$ in Hilbert space with $\|\psi\|=1$; by setting the joint probability distribution density for the first $n$ flashes at space-time points $(\vx_1,t_1), \ldots, (\vx_n,t_n)$ equal to
\begin{equation}
  \PPP_n^\psi(\vx_1,t_1, \ldots, \vx_n,t_n) =
  \Bigl\| \Lambda(\vx_n)^{1/2} \, W_{t_n -t_{n-1}} \cdots 
  \Lambda(\vx_1)^{1/2} \, W_{t_1-t_0} \psi \Bigr\|^2
\end{equation}
with $W_t = \exp\bigl(-\tfrac{i}{\hbar} Ht -\tfrac{1}{2} \int d^3 \vx \, \Lambda(\vx) \,t \bigr)$ for $t\geq 0$ and $W_t =0$ for $t<0$.

\begin{prop}\label{grw}
Let $\obs$ be a self-adjoint operator with pure point spectrum. In GRW-type theories with flash ontology as defined above, a strong superselection rule holds for $\obs$ if
\begin{subequations}\label{grwcond}
\begin{align}
  [\obs,\Lambda(\vx)] &=0 \quad \forall \vx \in \RRR^3,\\
  [\obs,H] &= 0\,.
\end{align}
\end{subequations}
\end{prop}

As an example, consider the quantum field theory from the beginning of Section~\ref{sec:ex}, for $\obs$ the total fermion number operator, and for $\Lambda(\vx)$ the (fermion + boson) particle number density operators, smeared out by convolution with a Gaussian. Then $[\obs,H] = 0$ and $[\obs,\Lambda(\vx)]=0$ (since all number operators commute with each other), so that Proposition~\ref{grw} applies. As another example, consider the example from the last paragraph of Section~\ref{sec:ex} supposing that 3-space has two connected components $C_1$ and $C_2$ due to nontrivial cosmology. Take $\obs$ again to be the number of the component containing particle 1 and $\Lambda(\vx)$ \z{the} multiplication operator by a Gaussian centered at $\vx$, which we take to be zero on $C_2$ if $\vx \in C_1$ and vice versa. Then \eqref{grwcond} is satisfied. As further examples, consider the two examples of Section~\ref{sec:unmeasurable}: a world without magnetic fields [or with magnetic fields only in the $z$ direction], and $\obs$ any spin component operator [respectively the $z$ component]. Given that $\Lambda(\vx)$ is the multiplication operator by a Gaussian centered at $\vx$ (times the identity in spin space), these examples are now cases of strong (instead of weak) superselection, since $\obs$ commutes with $\Lambda(\vx)$.

This situation in GRWf should be contrasted with the alternative version ``GRWm'' \cite{BG03,AGTZ} that is based on the matter density ontology. In this version, the primitive ontology is a continuous distribution of matter in 3-space with density given by
\begin{equation}
  m(\vx,t) = \sp{\psi_t}{\Lambda(\vx)|\psi_t}\,,
\end{equation}
where
\begin{equation}
  \psi_t = \frac{ W_{t-t_n} \Lambda(\vx_n)^{1/2}\, W_{t_n-t_{n-1}} 
  \cdots \Lambda(\vx_1)^{1/2}\, W_{t_1-t_0} \psi}{ \|W_{t-t_n} 
  \Lambda(\vx_n)^{1/2}\, W_{t_n-t_{n-1}} \cdots \Lambda(\vx_1)^{1/2}\, 
  W_{t_1-t_0} \psi\|}
\end{equation}
if $n$ collapses occurred between $t_0$ and $t$ and were centered at $(\vx_1,t_1), \ldots, (\vx_n,t_n)$. The
path space is replaced by a space $\histories$ of real-valued functions $m(\cdot)$ on space-time. Each element  $m(\cdot)$ represents the matter density of a certain history, and again, the initial wave function $\psi$ determines the probability distribution $\PPP^\psi$ on $\histories$, this time the probability distribution of the random function $m(\cdot)$. Thus, again, the definition of strong superselection around \eqref{sssr} is meaningful. However, in GRWm strong superselection presumably never holds. To see why, consider again the example in which 3-space $C_1 \cup C_2$ is not connected: During the very short period before the first collapse, both parts $\pvm_\obs(\ev) \psi$ of the wave function contribute to the matter density. This becomes particularly clear when there is only one particle, $N=1$: then, for every eigenfunction $\psi^\ev$, $m^{\psi^\ev}(\cdot)$ vanishes on one component of 3-space, but $m^\psi(\cdot)$ typically does not before the first collapse. An essential difference here between the GRWm and GRWf versions is that in the GRWm model the matter is supposed to exist for all times, while in the GRWf model space is empty at almost every time, and it is only at the instants of collapses that matter exists---in the form of flashes.

Concerning weak superselection, the concept is, of course, meaningful as well in the GRW model, and the model allows, like Bohm--Bell theories, to \emph{derive} which operators are observables. Since GRWm is empirically equivalent to GRWf \cite{AGTZ}, weak superselection holds in GRWm whenever it holds in GRWf, and thus in particular when strong superselection holds in GRWf, in particular under condition \eqref{grwcond}. This includes all examples mentioned above as examples of strong superselection in GRWf.


\section{Conclusions}
\label{sec:conclusions}

We have formulated two clear senses in which an operator $\obs$ can fulfill a superselection rule; there may exist further senses, perhaps more vague ones. The stronger sense that we have defined is grounded in the ``primitive ontology'': the particle trajectories in Bohm--Bell theories (and the flashes or matter density in the GRW model, see Section~\ref{sec:grw}). In particular, whether or not a strong superselection rule holds depends on the choice of the primitive ontology. The weaker sense is grounded in the impossibility of experimental distinction between $\psi$ and $\mu^\psi$, and thus in vague notions such as ``we'', ``macroscopic'', and ``observe''. Still, when these notions get based in turn on a clear primitive ontology, one can \emph{prove} weak superselection under suitable conditions.

For both weak and strong superselection of an operator $\obs$ we have formulated precise conditions on $\obs$, the Hamiltonians, and the configuration observable. One of our criteria implies that every joint symmetry of the Hamiltonian and the configuration observable, such as a gauge symmetry, gives rise to a weak superselection rule.

Conversely, an empirically obtained weak superselection rule for the operator $\obs$ can suggest a choice between several possible configuration observables, in two ways: Firstly, some choices may violate the weak superselection of $\obs$ (the relevant condition is presumably \eqref{GcommP}); thus, the weak superselection of $\obs$ can be an easy test of the empirical adequacy of a given Bohm--Bell model. Secondly, some choices may imply weak but not strong superselection of $\obs$; to the extent that one thinks of the superselection rule for $\obs$ as not merely apparent but fundamental, these choices appear less plausible.

\section*{Appendix}
\label{sec:proof}

\noindent \textit{Proof of Proposition 1.}
We begin by proving the statements (a)--(f) of Section~\ref{sec:cond}.

\begin{itemize}
\item[(a)] 
Since the Hamiltonian is assumed to be of the form $H= H_0 + H_I$ with $H_0$ given by \eqref{H0Schr} or \eqref{H0Dirac} and $H_I$ an integral operator, i.e.,
\begin{equation}
  H_I \psi(q) = \int_\Q dq' \, K(q,q') \, \psi(q')\,,
\end{equation}
one computes that the commutator of the Hamiltonian with
\begin{equation}\label{obsfctpvm}
  \obs = \int_\Q \fct(q) \, \pvm(dq)
\end{equation}
is given by
\begin{equation}\label{Hobscomm}
\begin{split}
  [H,\obs]\psi(q) =& \sum_i b_i(q) \frac{\partial \fct}{\partial q_i} \, \psi(q) + \sum_{i,j} 
  a_{ij}(q) \frac{\partial^2 \fct}{\partial q_i \partial q_j} \, \psi(q) +
  \sum_{i,j} 2a_{ij}(q) \frac{\partial \fct}{\partial q_i} \frac{\partial \psi}{\partial q_j}
  \\
  &+ \int_\Q dq' \, K(q,q') \, \bigl(\fct(q')-\fct(q) \bigr) \, \psi(q') = 0 \,.
\end{split}
\end{equation}
This implies $K(q,q')=0$ whenever $\fct(q') \neq \fct(q)$ because we can choose $\psi$ so that it vanishes identically outside an arbitrarily small neighborhood of $q'$ (not containing $q$, \z{so that the first three terms on the right hand side do not contribute). Therefore}, $[H_I,\obs] = 0$, and thus also $[H_0,\obs] =0$.

\item[(b)]
Observe from \eqref{obsfctpvm} that the spectral decomposition of the self-adjoint operator $\obs$ corresponds to the PVM on $\RRR$ acting on $\Hilbert$ given by
\begin{equation}\label{pvmobs}
  \pvm_\obs (\cdot) = \pvm(\fct^{-1}(\cdot))\,.
\end{equation}
This implies that the eigenvalues of $\obs$ are those $\ev \in \RRR$ for which $\fct^{-1}(\ev)$ is not a $\pvm$-null set. Since changes of $\fct$ on $\pvm$-null sets do not change $\obs$, we can choose $\fct$ so that it assumes only eigenvalues of $\obs$.

\item[(c)]
Since $[H_0,\obs] =0$ and this commutator is given explicitly by the first line of \eqref{Hobscomm}, one can read off that
\begin{equation}\label{nablafct}
  \nabla \fct =0\,.
\end{equation}
Indeed, for $H_0$ of the Schr\"odinger type \eqref{H0Schr} we can choose a $\psi$ with $\psi(q) =0$ and $\nabla \psi(q)$ any desired (complex) vector. Since the matrix $a_{ij}$ is of full rank, $\nabla \fct$ must vanish at $q$. For $H_0$ of the Dirac type \eqref{H0Dirac}, we can choose for any desired \z{direction $\boldsymbol{n}= (n_i)$ in configuration space} a $\psi$ such that $\sum_i n_i b_i(q) \psi(q) \neq 0$ since $\sum_i n_i b_i(q)$ is a matrix of full rank; thus the derivative of $\fct$ in the direction $\boldsymbol{n}$ must vanish.

From \eqref{nablafct} it follows that $\fct$ is
constant over every connected component of the configuration space $\Q$.

\item[(d)]
Note that $\fct(Q_t)$ could change with time in two ways: by continuous motion of $Q_t$, or by a jump. By continuous motion $Q_t$ cannot leave a connected component of $\Q$, on which, however, $\fct$ is constant. Alternatively and more directly from $[H_0,\obs] =0$, we can compute that
\begin{equation}
  \frac{d\fct(Q_t)}{dt} = v^{\psi} \cdot \nabla \fct (Q_t) = \Re
  \frac{\sp{\psi} { \pvm(dq) \tfrac{i}{\hbar} 
  [H_0, \obs] |\psi }}{\sp{\psi| \pvm(dq)}{\psi} } \bigg|_{q=Q_t} = 0\,.
\end{equation}
Let us next consider the stochastic jumps. The claim is that those jumps $q' \to q$ that would change the value of $\fct$, i.e., those with $\fct(q) \neq \fct(q')$, \z{have zero rate and thus} do not occur. To see this, recall that the spectral PVM of $\obs$ is given by \eqref{pvmobs}. Now let $B,B' \subseteq \RRR$ be disjoint intervals \z{with $\fct(q) \in B$ and $\fct(q')\in B'$}; then $\pvm_\obs(B) H_I \pvm_\obs(B') = 0$ because every spectral projection $\pvm_\obs(B)$ of $\obs$ commutes with $H_I$, and $\pvm_\obs (B) \pvm_\obs(B') = 0$ by the disjointness. \z{Thus,} 
\begin{equation}
  \sigma(dq|q') = \frac{ [\tfrac{2}{\hbar} \, \Im \, \sp{\psi}
  { \pvm(dq) H_I \pvm(dq') | \psi}]^+} {\sp{\psi} {\pvm(dq') | \psi}} = 0\,,
\end{equation}
which is what we wanted to show.

\item[(e)]
To arrive at \eqref{Pevpsi}, we need to check that the velocity $v^\psi$ and the jump rate $\sigma^\psi$ do not change when we replace $\psi$ by $\psi^\ev = N \, \pvm_\obs(\ev) \psi$ with normalizing constant $N= 1/\|\pvm_\obs(\ev) \psi\|$. Note first that $\pvm_\obs(\ev) = \pvm(\fct^{-1}(\ev))$ and thus, if $\fct(q) = \ev$, $\pvm_\obs(\ev) \pvm(dq) = \pvm(dq)$. Therefore, $\langle \psi^\ev| \pvm(dq) = N \, \langle \psi| \pvm(dq)$ and $\sp{\psi^\ev} {\pvm(dq) |\psi^\ev} = N^2 \, \sp{\psi} {\pvm(dq)| \psi}$. Since the operator $F$ in \eqref{bohm} and $\pvm_\obs(\ev)$ are both functions of $\pvm$, they commute; by (a), also $H_0$ and $\pvm_\obs(\ev)$ commute. Therefore, $\sp{\psi^\ev} {\pvm(dq) \tfrac{i}{\hbar} [H_0,F]| \psi^\ev} = N^2 \, \sp{\psi} {\pvm(dq) \tfrac{i}{\hbar} [H_0,F]| \psi }$, and so $v^{\psi^\ev} = v^\psi$. 

Similarly, if $\fct(q) = \ev = \fct(q')$, $\pvm(dq') \pvm_\obs(\ev) = \pvm(dq')$. Therefore 
\begin{equation}
  \sp{\psi^\ev}{\pvm(dq) H_I \pvm (dq') |\psi^\ev} = N^2 \, 
  \sp{\psi}{\pvm(dq) H_I \pvm(dq')| \psi}\,,
\end{equation}
and so $\sigma^{\psi^\ev} = \sigma^\psi$.

\item[(f)]
To arrive at \eqref{probpsiev}, simply observe that 
\[
  \PPP^{\psi^\ev}\Bigl(Q_t \in dq \Bigr) = \sp{\psi^\ev}
  { \pvm(dq) |\psi^\ev} = 1_{\{\fct = \ev\}}(q) \, N^2 \, 
  \sp{\psi}{\pvm(dq) | \psi}
\]
and
\[
  \PPP^\psi \Bigl(Q_t \in dq \Big| \fct(Q_t) = \ev \Bigr) 
  = \frac{\sp{\psi}{ \pvm(dq \cap \fct^{-1}(\ev))| \psi} }
  {\sp{\psi}{\pvm(\fct^{-1}(\ev)) | \psi}}\,,
\]
which is the same because
\[
  \pvm(dq \cap \fct^{-1}(\ev)) = 1_{\{\fct = \ev\}}(q) \, 
  \pvm(dq)\text{ and }N = \sp{\psi}{\pvm(\fct^{-1}(\ev)) | \psi}^{-1/2}\,.
\]
\end{itemize}

\noindent Now \eqref{sssr} follows from \eqref{Pevpsi} and \eqref{probpsiev} by considering for $\mu^\psi$ the mixture \eqref{mixture}.
\Endproof

\bigskip

\noindent \textit{Proof of Proposition~\ref{twoiffeight}.}
If $\fct(Q_t)$ is almost surely time-independent then its expectation 
\begin{equation}
  \EEE \fct(Q_t) = \int_\Q \fct(q) \, \sp{\psi_t} 
  {\pvm(dq)|\psi_t} = \sp{\psi_t}{G|\psi_t}
\end{equation}
is time-independent, too. Thus, 
\begin{equation}
  0= \frac{d}{dt} \sp{\psi_t}{\obs|\psi_t} = \sp{\psi_t}{\tfrac{i}{\hbar}[H,\obs]|\psi_t}
\end{equation}
for arbitrary initial $\psi$, and so $[H,\obs]=0$. The converse implication was established above under (d).
\Endproof

\bigskip

\noindent \textit{Proof of Proposition~\ref{propdet}.}
If $\sigma^\psi(dq|q') = 0$, and thus (replacing if necessary $q \leftrightarrow q'$) $\Im \sp{\psi}{\pvm(dq) H_I \pvm(dq')|\psi} = 0$ for all $\psi$, then $\pvm(dq) H_I \pvm(dq')$ is Hermitian. Thus, for $B,B' \subseteq \Q$, $\pvm(B) H_I \pvm(B') = \pvm(B') H_I \pvm(B)$, and for $B' = \Q$ we obtain $[H_I, \pvm(B)] =0$. Conversely, if $[H_I, \pvm(B)] = 0$ then we have for disjoint volume elements $dq$ and $dq'$ that $\sp{\psi}{\pvm(dq) H_I \pvm(dq')|\psi} = \sp{\psi}{\pvm(dq) \pvm(dq') H_I| \psi} = 0$ and thus $\sigma^\psi(dq|q') = 0$.
\Endproof

\bigskip

\noindent \textit{Proof of Proposition~\ref{propweak}.}
To begin with, let us make explicit that along with the decomposition of the Hilbert space, $\Hilbert = \Hilbert_\sys \otimes \Hilbert_\env$ we also intend that the configuration observable is formed from the configuration observables of $\sys$ and $\env$, 
\begin{equation}\label{pvmcompo}
  \pvm(B_\sys \times B_\env) = \pvm_\sys(B_\sys) 
  \otimes \pvm_\env(B_\env)\,. 
\end{equation}
We first show that \eqref{condwssr} implies that only those self-adjoint operators $A_\sys$ on $\Hilbert_\sys$ can correspond to observables that satisfy $[\obs, A_\sys] = 0$.  We do this by showing that for any state vector $\psi$ of $\sys$, the vector $\exp(i\obs s)\psi$ with any $s\in \RRR$ leads to the same probability distribution as $\psi$ of the result for every executable experiment, which implies that operators $A_\sys$ that do not commute with $\obs$ cannot ``be measured.''

To see this, suppose that the composite $\sys + \env$ starts with state vector $\Psi_0(s) = \exp(i\obs s) \psi \otimes \phi$ for some $\phi \in \Hilbert_\env$ and evolves during the experiment to $\Psi_t(s) = \exp(iHt)\Psi_0(s)$. If $B_\env^{(\alpha)}$ denotes the region in the configuration space of $\env$ in which the meter displays the outcome $\alpha$ of the experiment, the probability of the outcome $\alpha$ is 
\begin{equation}
  \prob_s(\alpha) = \sp{\Psi_t(s)}{[1_\sys \otimes \pvm_\env 
  (B_\env^{(\alpha)}) ] |\Psi_t(s)}\,. 
\end{equation}
By \eqref{condwssr}, $\obs \otimes 1_\env$ commutes with $H$ given by \eqref{Hcompo}, and thus $\Psi_t(s) = [\exp(i\obs s) \otimes 1_\env]  \Psi_t(0)$. Therefore, 
\[
  \prob_s(\alpha) 
  = \sp{\Psi_t(0)}{ [\exp(-i\obs s) \otimes 1_\env] [1_\sys \otimes 
  \pvm_\env(B_\env^{(\alpha)})] [\exp(i\obs s) \otimes 1_\env]| 
  \Psi_t(0)} = 
\]
\[
  = \sp{\Psi_t(0)}{[1_\sys \otimes \pvm_\env(B_\env^{(\alpha)}) ]
  | \Psi_t(0)} = \prob_0(\alpha) \,,
\]
independently of $s$, which is what we wanted to show.

As a consequence, a weak superselection rule holds for $\obs$: a state vector $\psi \in \Hilbert_\sys$ is empirically indistinguishable from the statistical mixture $\mu^\psi$ given by \eqref{mixture}. Indeed, we have seen that $\psi$ is empirically indistinguishable from $\exp(i\obs s)\psi$ for any $s \in \RRR$, and thus from any mixture of these, e.g., with $s$ uniformly distributed in $[0,S]$. But in the limit $S \to \infty$, the density matrix of this mixture converges, by standard decoherence theory, according to
\begin{equation}\label{rhoconvergence}
  \rho_S = \frac{1}{S} \int\limits_0^S ds\, e^{i\obs s} |\psi\rangle \langle \psi| e^{-i\obs s} 
  \xrightarrow{S\to \infty} \sum_\ev \pvm_\obs(\ev) |\psi\rangle
  \langle \psi| \pvm_\obs(\ev) = \rho^\psi\,,
\end{equation}
which is the density matrix \eqref{dm} of the mixture \eqref{mixture}. Since ensembles with the same density matrix are empirically indistinguishable, so are $\psi$ and $\mu^\psi$.

As an alternative argument, suppose an experiment corresponding to an observable $A=A_\sys$  is performed on ``$\sys$'', and suppose for simplicity that $A$ has pure point spectrum, so that its spectral decomposition is $A=\sum_\alpha \alpha \pvm_A(\alpha)$ with spectral projections $\pvm_A(\alpha)$. Then the probability of result $\alpha$ is $\sp{\psi}{\pvm_A(\alpha)|\psi}$, which coincides with the probability of result $\alpha$ from the mixture \eqref{mixture}, $\tr (\pvm_A(\alpha) \rho^\psi) = \sum_\ev \sp{\psi} {\pvm_\obs(\ev) \pvm_A(\alpha) \pvm_\obs(\ev) |\psi}$, because $\pvm_A(\alpha)$ and $\pvm_\obs(\ev)$ commute.
\Endproof

\bigskip

\noindent \textit{Proof of Proposition~\ref{GHPweak}.}
By $[\obs,H]=0$, the mixture $\mu^\psi$ given by \eqref{mixture} evolves during $t$ units of time into $\mu^{\psi_t}$. By $[\obs,\pvm(dq)]=0$ and thus $[\pvm_\obs(\ev), \pvm(dq)]=0$, $\mu^{\psi_t}$ yields the same distribution of the configuration as $\psi_t$, namely
\[
  \int\limits_{\sphere(\Hilbert)} \mu^{\psi_t}(d\phi) \, \sp{\phi}
  {\pvm(dq) | \phi} = \tr \bigl(\pvm(dq) \, \rho^{\psi_t} \bigr) =
\]
\[
  = \sum_\ev \sp{\psi_t} {\pvm_\obs(\ev) \, \pvm(dq) \, \pvm_\obs(\ev)| \psi_t} =
  \sp{\psi_t} {\pvm(dq)| \psi_t}
\]
using \eqref{dm}. Since the outcome of any experiment is read off from the configuration $Q_t$ at some time $t$, the fact that the distribution of $Q_t$ is the same for $\psi$ and $\mu^\psi$ implies that the distribution of the outcome is the same for the two, so that the experiment cannot distinguish the two.
\Endproof

\bigskip

\noindent \textit{Proof of Proposition~\ref{grw}.}
The distribution of the first $n$ flashes arising from the mixture \eqref{mixture} is
\begin{multline}
  \sum_\ev \|\pvm_\obs(\ev)\psi\|^2 \, \PPP_n^{\psi^\ev} 
  (\vx_1,t_1, \ldots, \vx_n,t_n) =\\ \sum_\ev \sp{\pvm_\obs(\ev) \psi} 
  {W_{t_1 -t_{0}} \Lambda(\vx_1)^{1/2} \cdots 
  \Lambda(\vx_1)^{1/2} W_{t_1-t_0} |\pvm_\obs(\ev) \psi}\,.
\end{multline}
Since, by \eqref{grwcond}, $[\obs,W_t] =0$ and $[\obs,\Lambda(\vx)^{1/2}]=0$, and therefore $[\pvm_\obs(\ev),W_t]=0$ and $[\pvm_\obs(\ev),\Lambda(\vx)^{1/2}]=0$, this quantity equals
\begin{multline}
  \sum_\ev\sp{\psi} 
  {\pvm_\obs(\ev) W_{t_1 -t_{0}} \Lambda(\vx_1)^{1/2} \cdots 
  \Lambda(\vx_1)^{1/2} W_{t_1-t_0} | \psi} =\\
  \sp{\psi} 
  {W_{t_1 -t_{0}} \Lambda(\vx_1)^{1/2} \cdots 
  \Lambda(\vx_1)^{1/2} W_{t_1-t_0} |\psi} = \PPP^\psi_n (\vx_1,t_1, \ldots, 
  \vx_n,t_n)\,,
\end{multline}
the distribution arising from $\psi$.
\Endproof

\end{document}